\documentstyle[12pt]{article}
\begin{document}
\begin{titlepage}
\hspace*{\fill} ULB--TH--94/18\\
\hspace*{\fill} NIKHEF--H--94--34\\
\hspace*{\fill} KUL--TF--94--37\\
\hspace*{\fill} hep-th/9411202\\
\vspace{0.5cm}

\begin{centering}

{\huge Conserved currents and gauge invariance in Yang-Mills theory}

\vspace{1cm}
{\large Glenn Barnich$^{1,*}$, Friedemann Brandt $^{2,**}$ and \\
Marc Henneaux$^{1,***}$}\\
\vspace{1cm}
$^1$Facult\'e des Sciences, Universit\'e Libre de Bruxelles,\\
Campus Plaine C.P. 231, B-1050 Bruxelles, Belgium\\
$^2$NIKHEF-H, Postbus 41882, 1009 DB Amsterdam, The
Netherlands;\\  after November 1, 1994 : Instituut voor Theoretische
Fysica, K.U. Leuven,\\
Celestijnenlaan 200D, B-3001 Leuven, Belgium\\
\vspace{0.5cm}

\begin{abstract}
It is shown that in the absence of free
abelian gauge fields,
the conserved currents of (classical) Yang-Mills gauge models
coupled to matter fields can be always redefined so as to be gauge
invariant.  This is a direct consequence of the general
analysis of the Wess-Zumino consistency condition for
Yang-Mills theory that we have provided recently.
\end{abstract}
\vspace{0.5cm}

{To appear in {\it Phys.~Lett.~B}}\\
\end{centering}

\vspace{0.5cm}

{\footnotesize \hspace{-0.6cm}($^*$)Aspirant au Fonds National de la
Recherche
Scientifique (Belgium).\\
($^{**}$)Supported by Deutsche Forschungsgemeinschaft and
by the research council (DOC) of the K.U. Leuven under
Grant No. F94/22.\\
($^{***}$)Also at Centro de Estudios
Cient\'\i ficos de Santiago, Chile.}
\end{titlepage}

\pagebreak

\def\qed{\hbox{${\vcenter{\vbox{
   \hrule height 0.4pt\hbox{\vrule width 0.4pt height 6pt
   \kern5pt\vrule width 0.4pt}\hrule height 0.4pt}}}$}}
\newtheorem{theorem}{Theorem}
\newtheorem{lemma}{Lemma}
\newtheorem{definition}{Definition}
\newtheorem{corollary}{Corollary}
\newcommand{\proof}[1]{{\bf Proof.} #1~$\qed$.}

It is well known that the standard Noether method
\cite{Noether} for deriving the
energy momentum tensor of the Yang-Mills field from the invariance
of the Yang-Mills action under space-time translations does not yield a
gauge invariant answer if one regards the connection $A^a_\mu$ as an
ordinary covector,
\begin{eqnarray}
\delta_\xi A^a_\mu={\cal L}_\xi A^a_\mu=a^\rho\partial_\rho A^a_\mu
\label{1}
\end{eqnarray}
($\xi=a^\rho\partial /\partial x^\rho$). Rather, one gets the
Yang-Mills canonical
energy momentum tensor
\begin{eqnarray}
\Theta^\lambda_\mu=F^{\lambda\rho}_a\partial_\mu A^a_\rho-{1\over 4}
\delta^\lambda_\mu F^{\rho\sigma}_a F^a_{\rho\sigma},\label{2}
\end{eqnarray}
which fails to be gauge invariant
because of the first term. It has been
observed by Jackiw \cite{Jackiw}, however, that if one ``improves" the
transformation law (\ref{1}) by adding to it an appropriately
chosen gauge transformation,
\begin{eqnarray}
\delta_\xi A^a_\mu\longrightarrow \delta^{imp}_\xi A^a_\mu
={\cal L}_\xi A^a_\mu+D_\mu(-A^a_\rho a^\rho)
=a^\rho F^a_{\rho\mu},\label{3}
\end{eqnarray}
then, the standard Noether procedure yields the
Yang-Mills energy momentum tensor
\begin{eqnarray}
T^\lambda_\mu=F^{\lambda\rho}_a F^a_{\mu\rho}-{1\over 4}
\delta^\lambda_\mu F^{\rho\sigma}_a F^a_{\rho\sigma},\label{4}
\end{eqnarray}
which is gauge invariant.  The ``improved"
diffeomorphisms (\ref{3}) for Yang-Mills gauge
fields have been discussed by various
authors in different contexts,
see e.g. \cite{Teitelboim}.
The conserved quantities (\ref{2})
and (\ref{4}) differ on-shell
by an identically conserved current and yield the same integrated
conserved charges. However, it
is only (\ref{4}) that is gauge invariant and
that describes correctly the local
coupling of the Yang-Mills field to the
gravitational field.

In fact one can improve analogously all infinitesimal conformal
transformations of the $A_\mu^a$ so as to get
corresponding gauge invariant conserved Noether
currents \cite{Jackiw}.
This  raises
immediately the following question : can the same
``improvement" be achieved for
{\it any} global symmetry of the
Yang-Mills theory ? We show in this
letter that the answer is affirmative
provided that there is no free abelian
gauge field. Our results are in
fact direct consequences of the analysis
of the Wess-Zumino consistency condition that we have carried out in
\cite{BBH1,BBH2}. For this reason,
we shall closely follow the language
and the notations of these references.  In particular, we consider
only polynomial functions of the fields, the antifields and
their derivatives, and we take the spacetime dimension
$n$ to be greater than or equal to $3$.

The question of the gauge invariance of
the conserved currents has indirectly arisen
recently in the very interesting work
\cite{Torre}, where the symmetries
of the Yang-Mills equations have been all
systematically derived under simplifying
assumptions, one of which was the
gauge covariance of the symmetry transformation
laws. The analysis of this paper
establishes that this assumption is really not necessary
for symmetries of the Yang-Mills action,
since it can always be fulfilled by redefinitions.
Earlier insights into the structure of the conserved
currents for Yang-Mills models may be found in \cite{DesNic}.

That the conserved currents
can be chosen to be gauge invariant
is perhaps not surprising
since there are probably no non trivial
conserved currents besides those associated with the known global
symmetries.  These can be chosen to be
gauge invariant.  However, to our knowledge, that result has never
been proved completely before.

We shall assume throughout the
letter that the Yang-Mills structure
Lie algebra is the direct sum of
a semi-simple Lie algebra plus abelian
ideals (``reductive Lie algebra").
We allow for (minimal) couplings to spin $0$ and spin $1/2$ matter fields
transforming under linear
representations of the gauge group.  Thus, our
analysis covers e.g. globally supersymmetric
Yang-Mills models.
However, we exclude matter fields
carrying a gauge invariance of their own.
Remarks concerning the gravitational
case are given at the end.

We shall first consider the
case when the abelian
gauge fields (if any) are coupled to at least one charged
matter field.  The theorem holds only under
that physically natural assumption.  We shall next indicate explicitly
what happens in the presence of free abelian gauge
fields, for which we shall
give the complete list of all the currents that cannot be
covariantized (in $n \not=4$ dimensions).

As it is usual in the physics literature, we call ``symmetry" of the
theory a transformation of the fields
\begin{eqnarray}
\Delta\phi^i=a^i(x,\phi,\partial\phi,\dots,\partial^k\phi)
\end{eqnarray}
($(\phi^i)\equiv(A^a_\mu,matter\ fields)$) which leaves the Lagrangian
${\cal L}$ invariant up to a total derivative,
\begin{eqnarray}
{\delta {\cal L}\over\delta \phi^i}\Delta\phi^i+\partial_\mu j^\mu=0
\label{6}
\end{eqnarray}
(``variational symmetry" in the mathematics literature). If one introduces
the antifields $\phi^*_i$ associated with the fields $\phi^i$, the
equation (\ref{6}) may be rewritten as
\begin{eqnarray}
\delta a_1 +\partial_\mu j^\mu=0 . \label{7}
\end{eqnarray}
Here, $a_1\equiv \phi^*_i a^i$ and $\delta$ is the field theoretical
Koszul-Tate differential \cite{Fisch}. Explicitly,
\begin{eqnarray}
\delta \phi^*_i={\delta^R {\cal L}\over\delta \phi^i},\qquad
\delta \phi^i=0\cr
\delta C^*_a=-D_\mu A^{*\mu}_a+gT^j_{ai}y^*_jy^i,\qquad \delta C^a=0
\label{delta}
\end{eqnarray}
where the $T^j_{ai}$ denote a basis of
generators for the linear representation of the
gauge group under which the matter fields $y^i$ transform.
As usual, the $C^a$ in (\ref{delta}) stand for the ghosts.  They
have their own antifields, denoted by $C^*_a$.  The Koszul-Tate differential
implements the equations of motion in cohomology, i.e., any on-shell
vanishing function may be written as a $\delta$-variation.  Equ.
(\ref{7}) expresses simply that the divergence of the current $j^\mu$
is zero on-shell.

It follows from (\ref{7}) that a symmetry defines
an element of the cohomological group
$H^n_1(\delta|d)$ \cite{BBH1}, where $n$ is the form degree and $1$ is
the antighost number (we are using dual notations;
in form-notations, $a_1$ would be a
$n$-form and $j^\mu$ a $(n-1)$-form).

A symmetry is said to be a trivial
global symmetry if the corresponding
$a_1$ is in the trivial class of
$H^n_1(\delta|d)$, i.e., if it can be
written as
\begin{eqnarray}
a_1=\delta b_2+\partial_\mu c^\mu\qquad (trivial\ a_1)\label{8}
\end{eqnarray}
for some $b_2$ and $c^\mu$. This happens
if and only if the transformation
$a^i$ reduces on-shell to a gauge
transformation.  That $a^i$ reduces
on-shell to a gauge symmetry when (\ref{8}) holds is obvious.
That, conversely, any symmetry transformation that reduces on-shell
to a gauge symmetry can be written as in
(\ref{8}) follows from the facts
(i) that gauge symmetries are of the
form (\ref{8})~; and (ii) that any symmetry
transformation leaving the Lagrangian
invariant up to a total derivative and vanishing on-shell is
necessarily an antisymmetric combination of the
field equations and thus of
the form (\ref{8}) (see \cite{BBH1} sections 7 and 11).

The above redefinition (\ref{3}) of the translations is of this type,
i.e. the solutions $a_1$ and $a_1^{imp}$ of (\ref{7}) corresponding
to (\ref{1}) and (\ref{3}) satisfy $a_1-a_1^{imp}=
\delta b_2+\partial_\mu c^\mu$ for some $b_2$ and $c^\mu$.
For a trivial global symmetry, the conserved
current identically fulfills
$\partial_\mu (j^\mu+\delta c^\mu)=0$ from (\ref{7}). Thus,
\begin{eqnarray}
j^\mu+\delta c^\mu=\partial_\nu S^{\nu\mu},
\quad S^{\nu\mu}=-S^{\mu\nu}
\qquad (trivial\quad j^\mu),  \label{trivial}
\end{eqnarray}
which implies that $j^\mu$ reduces
on-shell to an identically conserved current
and is ``trivial"\footnote{The terminology is borrowed
from the mathematical
literature and does not imply
that the currents are necessarily physically trivial.  The conserved
currents associated with gauge symmetries reduce indeed
to divergences \cite{Noether}
and the corresponding conserved charges are
given by non-vanishing physically relevant surface integrals.
Furthermore, on-shell trivial terms
must be handled with care inside Green functions
in the quantum theory.
For the purposes of this letter, however, it is
convenient to call a current of the form (\ref{trivial})
``trivial".}.
Conversely, if the current is trivial, then $a_1$ is
necessarily of the form (\ref{8}) \cite{BBH1}.
This means that a trivial
redefinition of the symmetry is completely equivalent to a trivial
redefinition of the current (see
\cite{Cangemi} for related considerations).

Now, the current is gauge invariant if
and only if it is annihilated by the
$\gamma$ piece of the BRST differential $s=\delta+\gamma$,
\begin{eqnarray}
\gamma j^\mu=0. \label{10}
\end{eqnarray}
Explicitly, $\gamma$ reads \cite{BRST},
\begin{eqnarray}
\gamma A^a_\mu=D_\mu C^a,\ \gamma C^a={1\over 2}g C^a_{bc}C^b C^c,
\ \gamma y^i=  g T^i_{aj}C^a y^j\nonumber\\
\gamma A^{*\mu}_a= g C^b C^c_{ab}A^{*\mu}_c,\ \gamma C^*_a=
 g  C^c_{ab}C^b C^*_c,\ \gamma y^*_i= -  g T^j_{ai} C^a y^*_j,
\label{11}
\end{eqnarray}
where $C^a_{bc}$ are the structure constants of the
Lie algebra. The differential $\gamma$ incorporates the
gauge symmetry since the $\gamma$-variations of the fields $A^a_\mu$
and $y^i$ are
obtained by
replacing the gauge
parameters by the ghosts in the gauge variations of the fields.
The $\gamma$-variations of the ghosts are such that $\gamma^2 = 0$.
The $\gamma$-variations of the antifields (sources for the BRST
variations of the fields) are also quite standard and express
the fact that the antifields $A^{*\mu}_a$ and $C^*_a$ transform in the
co-adjoint representation, while the antifields $y^*_i$ transform in
the representation dual to that of the $y$'s.  Finally, the
complete BRST differential $s$ takes both the
equations of motion (through
$\delta$) and the gauge symmetry (through $\gamma$)
into account.  It is
nilpotent because $\delta^2 =
\delta \gamma + \gamma \delta = \gamma^2 = 0$.

Condition (\ref{10}) implies
$\gamma\delta a_1=0$. Accordingly, a necessary
condition for the existence of trivial
redefinitions that make $j^\mu$
gauge invariant is that there
exist trivial redefinitions that make
$\delta a_1$ invariant. Our first
task will be to prove more, namely,
that one can always perform
redefinitions that make $a_1$ itself invariant,
$\gamma a_1=0$. The invariance of $a_1$ means that the infinitesimal
variations $\Delta\phi^i=a^i$ transform  in
representations contragredient to the
representations of the antifields.
Thus $\Delta A^a_\mu$ transforms in
the adjoint representation of the
gauge group (as it does in (\ref{3})
above), while $\Delta y^i$
transforms as $y^i$ for all matter fields.  This
implies that the global symmetry written
in its covariant version commutes with the
gauge transformations.

\begin{theorem}{\em{\bf :}}
In the absence of uncoupled abelian gauge fields,
any cohomological class of $H^n_1(\delta|d)$
contains a $\gamma$-invariant
representative. That is, by adding to any symmetry $a_1= \phi^*_i a^i$ a
$\delta$-trivial term modulo $d$,
\begin{eqnarray}
a_1\longrightarrow a^\prime_1 =a_1+\delta b_2 +\partial_\mu
c^\mu_1,
\end{eqnarray}
one can arrange so that the equivalent symmetry $a^\prime_1$
fulfills $\gamma a^\prime_1=0$.
\end{theorem}

\proof{
We start with the equation $\delta a_1+\partial_\mu j^\mu=0$.
Acting with $\gamma$ on it, one gets $\delta\gamma
a_1=\partial_\mu(\gamma j^\mu)$, i.e., $[\gamma a_1]\in
H^n_1(\delta|d)$. But $\gamma a_1$ involves the
ghosts and thus is $\delta$-exact modulo $d$
\cite{Henneaux},
\begin{eqnarray}
\gamma a_1=-\delta a_2+\partial_\mu j^{\prime\mu}.
\end{eqnarray}
By acting with $\gamma$ on this equation, one then finds that
$\gamma a_2$ is also $\delta$-trivial modulo $d$. This enables one to
construct recursively the sum
\begin{eqnarray}
a=a_1+a_2+\dots+a_k\label{14}
\end{eqnarray}
such that
\begin{eqnarray}
sa+\partial_\mu k^\mu=0  \label{16}
\end{eqnarray}
for some $k^\mu$. The construction, which follows the
standard lines of homological perturbation theory
\cite{Stasheff,Fisch} is reviewed in \cite{BBH1,BBH2}, where it is
explicitly demonstrated that the sum (\ref{14}) terminates after a
finite number of steps (\cite{BBH2}, section 3). Therefore, $a$
defines an element of $H(s|d)$.

The analysis of the cohomological groups $H(s|d)$ made in
\cite{BBH2} shows then that one can successively remove $a_k$,
$a_{k-1}$, $a_{k-2},\dots,a_2$ from $a$ by adding an
appropriate $s$-boundary modulo $d$.
At the same time, one may redefine $k^\mu$ so that it reduces
to a single term of antighost number zero, $k^\mu \rightarrow
k^{\prime \mu} = k^\mu + s t^\mu + \partial_\nu S^{\mu \nu}$,
$S^{\mu \nu} = - S^{\nu \mu}$, $k^{\prime \mu} =
k^{\prime \mu}_0$. After this is done, $a$
reduces to a term of antighost number $1$,
\begin{eqnarray}
a\longrightarrow a^\prime=a+sm+\partial_\mu n^\mu, \qquad
a^\prime=a^\prime_1\cr m=m_2+\dots m_{k+1},\qquad
n^\mu=n_1^\mu+\dots+n_k^\mu.
\end{eqnarray}
But adding to $a$ such a $s$-boundary modulo $d$ amounts to
adding to $a_1$ a $\delta$-boundary modulo $d$,
\begin{eqnarray}
a_1\longrightarrow a_1^\prime=a_1+\delta m_2+\partial_\mu
n_1^\mu,
\end{eqnarray}
Furthermore, the condition $sa^\prime+\partial_\mu
k^{\prime \mu} =0$
with $a^\prime=a^\prime_1$
and $k^{\prime \mu} =
k^{\prime \mu}_0$ implies precisely
\begin{eqnarray}
\delta a_1^\prime+\partial_\mu t^\mu_0=0
\end{eqnarray}
and
\begin{eqnarray}
\gamma a^\prime_1=0.
\end{eqnarray}
This proves the theorem.}

The removal of $a_k$, $a_{k-1}$, $a_{k-2},\dots,
a_2$ from $a$ by
adding $s$-coboundaries modulo $d$ can be performed because
the (invariant) cohomological groups $H_i(\delta|d)$ vanish for
$i\geq2$ \cite{BBH2}. If there were free abelian gauge fields,
which we denote by $A^\alpha_\mu$, then $H_2(\delta|d)$ would
not vanish. The best that can be
done then is to remove all the terms up
to $a_3$ included, $a\longrightarrow a^\prime=a^\prime_1+a^\prime_2$.
The term $a^\prime_2$ cannot be removed
in general and
is of the form worked
out in section 8 of \cite{BBH2} (``solution of class $I_b$").
The representatives of $H^{-1}(s|d)$ arising from
the non trivial elements of $H_2(\delta|d)$
fall into two categories : those that do not depend on
the spacetime coordinates, and those that do.
A complete list of
representatives not involving $x$ is given by
\begin{eqnarray}
a = a_1 + a_2 \label{solns}
\end{eqnarray}
with
\begin{eqnarray}
a_1=f_{\alpha\beta} A^\alpha_\mu A^{*\beta\mu}, \  a_2 = f_{\alpha\beta}
C^\alpha C^{*\beta}, \  f_{\alpha\beta}=-f_{\beta\alpha}. \label{solns2}
\end{eqnarray}
The corresponding symmetries are
\begin{eqnarray}
\Delta A^\alpha_\mu=f^\alpha_\beta A^\beta_\mu
\label{19}
\end{eqnarray}
($\Delta$ (other fields)$=0$) and simply rotate the free abelian fields
among themselves.  Here, we lower and
raise the indices $\alpha$, $\beta$, $\dots$
labeling the free abelian gauge fields with
the metric $\delta_{\alpha \beta}$ ($\delta^{\alpha \beta}$).
The conserved currents associated
with (\ref{19}) are
\begin{eqnarray}
j^\mu_{\alpha\beta}=F^{\mu\nu}_{[\alpha}A_{\beta]_\nu}
\end{eqnarray}
and cannot be redefined so as to be gauge invariant.
Hence, theorem 1 does indeed
not hold in the presence of free abelian gauge
fields.

We note that the solutions (\ref{solns}) have
counterparts with ghost number zero,
\begin{eqnarray}
a = a_0 + a_1 + a_2
\end{eqnarray}
with
\begin{eqnarray}
a_0 = \frac{1}{2} f_{[\alpha \beta \gamma]} A^\alpha_\mu
A^\beta_\nu F^{\nu \mu \gamma}
\end{eqnarray}
and
\begin{eqnarray}
a_1 = f_{[\alpha \beta \gamma]} C^\alpha A^\beta_\mu A^{*\mu \gamma},
\  a_2 = \frac{1}{2} f_{\alpha \beta \gamma}C^\alpha C^\beta C^{*\gamma},
\ f_{\alpha \beta \gamma} = f_{[\alpha \beta \gamma]}.
\end{eqnarray}
These give rise to the Yang-Mills cubic vertex and the non
abelian deformation of the gauge symmetries.

We now turn to the $x$-dependent solutions and consider only
the pure, free abelian case.  Preliminary results appear to
indicate that
in four dimensions, there are no $x$-dependent non trivial
solutions of (\ref{16}) whose part $a_2$ cannot
be removed.
By contrast, flat $n$-dimensional space-time admits for
$n \neq 4$ further
$x$-dependent solutions of (\ref{16}) with non trivial $a_2$,
for which it is accordingly impossible to
covariantize
the corresponding symmetries.
 They are given by
\begin{eqnarray}
\Delta A^\alpha_\mu=f^\alpha_\beta(x^\nu F^\beta_{\nu\mu}+{{n-4}\over2}
A^\beta_\mu),\qquad f_{\alpha\beta}=f_{\beta\alpha}\label{21}
\end{eqnarray}
with currents $j^\mu=f^\alpha_\beta j^{\;\; \beta \mu}_\alpha$
given by
\begin{eqnarray}
j^\mu= f^\alpha_\beta\left(F^{\mu\rho}_\alpha
F^\beta_{\nu\rho} x^\nu
-{1\over4} F^{\nu\rho}_\alpha F^\beta_{\nu\rho} x^\mu +
{{n-4}\over2}F^{\mu\rho}_\alpha
A^\beta_\rho \right).
\end{eqnarray}
The transformations (\ref{21}) contain pure scale transformations
($f^\alpha_\beta = \lambda \delta^\alpha_\beta$)
and exist in any number of dimensions, but it is only in
four dimensions that they admit a covariant expression.
Furthermore, they cease to be symmetries on
a generic curved background, in contrast to (\ref{19}).
The clash between conformal invariance and gauge invariance
for $p$-form gauge fields off their critical
dimension has been analysed in depth in \cite{Deser}.
We remark that given a symmetry for a model with free abelian gauge fields,
one can always subtract from it a symmetry of the type (\ref{19}) or
(\ref{21}) ($n \not= 4$) so that
$a_2$ can be made to vanish. The remaining symmetry
admits a gauge invariant conserved current.

We are now in a position to formulate
and prove the main theorem of this
letter~:

\begin{theorem}{\em{\bf :}}
In Yang-Mills theory without uncoupled
abelian gauge fields, one can always
redefine the conserved currents by
the addition of trivial terms,
\begin{eqnarray}
j^\mu\longrightarrow j^{\prime\mu}
\approx j^\mu+\partial_\nu S^{\nu\mu},
\ S^{\nu\mu}=-S^{\mu\nu},
\end{eqnarray}
in such a way that the equivalent
currents $j^{\prime\mu}$ are gauge invariant ($\approx$
denotes equality up to terms that vanish on-shell).
\end{theorem}

\proof{We have shown that the
symmetry tranformation may be assumed to be
gauge invariant,
\begin{eqnarray}
\delta a_1+\partial_\mu j^\mu=0\nonumber\cr
\gamma a_1=0.
\end{eqnarray}
But if $a_1$ - which does not
involve the ghosts - is annihilated by
$\gamma$, then it is an invariant polynomial in the field strengths,
the matter fields, the antifields and their covariant derivatives
(linear in $\phi^*_i$).
Then $\delta a_1$
itself is also an invariant polynomial
in the field strengths, the matter
fields and their covariant derivatives. This
invariant polynomial is a
divergence, and the question is whether the current $j^\mu$
of which it is the divergence may also
be assumed to be an invariant polynomial.
The obstructions to taking $j^\mu$
invariant have been studied in the literature
and are given by the invariant polynomials
in the undifferentiated curvature
$2$-forms $F^a={1\over2}F^a_{\mu\nu}dx^\mu dx^\nu$
\cite{Brandt,DuboisViolette}. [The invariant
cohomology of $d$ has been
studied for gravity in \cite{Gilkey} and recently from a general point
of view in \cite{Anderson}]. Since
the Yang-Mills equations involve the
differentiated curvatures, the
obstructions for taking $j^\mu$ invariant
are absent and thus, one may assume $\gamma j^\mu=0$.
This proves the theorem.}

Note that we have explicitly used here the fact
that the Lagrangian is the Yang-Mills one.
One can easily construct gauge
invariant
Lagrangians different from the
Yang-Mills one that possess covariant global
symmetries whose corresponding current is not gauge
invariant.
For example, the Lagrangian
\begin{eqnarray}
{\cal L}=\lambda tr F^2\label{x}
\end{eqnarray}
with Lagrange multiplier $\lambda$ leads to the equation $tr F^2=0$.
The characteristic polynomial
$tr F^2$ can be written as a divergence,
$tr F^2=dQ$, but one cannot choose the
conserved current $Q$ to be gauge
invariant, although the corresponding symmetry is covariant (it is
simply given by $\Delta\lambda=const$, $\Delta A_\mu^a=0$).

We conclude this letter by indicating
how our results extend to other theories
with a gauge freedom. The crucial tool by which we have controlled
the covariance of the symmetry
transformations and the invariance of the
currents is the vanishing of the (gauge invariant) cohomology groups
$H^n_i(\delta|d)$ for $i\geq 2$. These groups are
isomorphic to the groups
$H^{n-i}_0(d|\delta$) of the
characteristic cohomology of \cite{Bryant}.
It is thus the vanishing of the characteristic cohomology in form
degree $<n-1$ that controls the gauge invariance of the
currents for Yang-Mills models.
One expects similar results for
any theory in which $H^{n-i}_0(d|\delta)
\simeq H^n_i(\delta|d)=0$
for $i\geq 2$. And indeed, in the
case of Einstein gravity for which this
property holds, one may also choose
the conserved currents $j^\mu$ to
behave properly as vector densities.
This will be explicitly proved in a
separate publication \cite{BBH3}.
Similarly, if one modifies the Yang-Mills
action by adding terms that (i) preserve the
triviality of $H^n_i(\delta|d)=0$
for $i\geq 2$; and (ii) do not make any of the
characteristic classes vanish on-shell, then, one can
still assume that the conserved currents are
gauge invariant.

{\bf Acknowledgements}

We thank Ian Anderson and Stanley Deser for
useful conversations.
M.H. is grateful to the Institute for Advanced Study
(Princeton) for kind hospitality while this work was
being completed.
This research has been supported in part
by a research grant from the F.N.R.S.
(Belgium) and by research contracts with the Commission of the
European Community.

\vfill
\eject

\end{document}